\documentclass[%
 aip,
 amsmath,amssymb,
 reprint,%
]{revtex4-1}

\usepackage{graphicx}% Include figure files
\usepackage{dcolumn}% Align table columns on decimal point
\usepackage{bm}% bold math
\usepackage[utf8]{inputenc}
\usepackage[T1]{fontenc}
\usepackage{mathptmx}
\usepackage{wrapfig}
\usepackage{color}

\newcommand{\bu}{\mathbf u}
\newcommand{\bn}{\mathbf n}

\newcommand{\bx}{\mathbf x}
\newcommand{\bP}{\mathbf P}
\newcommand{\bI}{\mathbf I}
\newcommand{\bD}{\mathbf D}
\newcommand{\br}{\mathbf r}

\newcommand{\blf}{\mathbf f}

\newcommand{\Div}{\operatorname{div}}

\begin{document}

\preprint{AIP/123-QED}

\title[]{Speed--direction description of turbulent flows}
% Force line breaks with \\

\author{Maxim A. Olshanskii}%
 \email{molshan@math.uh.edu}
 %\homepage{www.math.uh.edu/~molshan}
\affiliation{
Department of Mathematics, University of Houston, Houston, TX, 77204.
}%

\date{\today}% It is always \today, today,
             %  but any date may be explicitly specified

\begin{abstract}
 In this note we  introduce  speed  and direction variables to describe the motion of incompressible viscous fluid. Fluid velocity $\bu$ is decomposed into  $\bu=u\br$, with $u=|\bu|$ and $\br=\bu/|\bu|$. We consider a directional split of the Navier--Stokes equations into a coupled system of equations for $u$ and  for $\br$. Equation for $u$ is particularly simple but solely maintains the energy balance of the system.
  Under the assumption of a weak correlation between fluctuations in speed and direction in a developed turbulent flow, we further illustrate the application of $u$-$\br$ variables to describe mean statistics of a shear turbulence. The standard (full) Reynolds stress tensor does not appear in a  resulting equation for the mean flow profile.
  \end{abstract}

\maketitle

\section{Introduction}
It is most common to describe flow phenomena in terms of fluid velocity, pressure and density, which are natural kinematic and thermodynamic variables. At the same time, the choice of alternative variables may give remarkable new insights. One prominent example is the employment of vorticity and streamfunction to devise a system of equations governing incompressible flow dynamics~\cite{majda2002vorticity}. { Application of these variables to study turbulent flows can be found, e.g. in ~\cite{yoon2016contribution,kareem2019hyperbolic,maulik2019subgrid} among many other publications. }
In this note, we consider the speed of the flow and its direction as independent Eulerian variables together with (kinematic) pressure to describe the motion of incompressible viscous  fluid. For fluid velocity field $\bu$, transformation to new variables formally takes the form $u=|\bu|$ and $\br=\bu/|\bu|$, where for $u=0$ the direction is ambiguously defined.  After a little calculus in section~\ref{sNSE}, the Navier--Stokes equations are written in terms of $u$, $\br$ and $p$. We next split the system by projecting the momentum equation  on the flow direction and the orthogonal plane in each point of space--time. The first scalar equation can be interpreted as the equation governing the evolution of $u$. This equation turns out to be particularly simple, but together with the incompressibility condition it encodes  an important physics in the form of energy balance.
Motivated by these observations, we further employ the new variables to describe turbulent flows. One key observation here is that speed and direction of velocity fluctuations, independent in isotropic turbulence, may still correlate  weakly in more practical flows.
More precisely, we  assume and check using DNS data for the channel turbulence that the second term in \eqref{v_decomp} can be neglected and so the mean velocity can be written in terms of mean speed and mean direction. { An elementary analysis reveals that the neglected term is also second order with respect to rms fluctuations in velocity field.} Mean speed $\overline{u}$ satisfies an equation, which results from averaging the Navier--Stokes equations projected on the flow direction; see eq.~\eqref{mean_prof}. It is interesting to see that the turbulent part appears in the equation in the form of correlation functions different from the well-known Reynolds stress.
The paper discusses and examines all terms in the equation for $\overline{u}$ for the example of a turbulent flow in a channel. For this purpose we make use of the turbulent channel flow data set from the Johns Hopkins turbulence database~\cite{kanov2015johns}.
In particular, a simple analytical representation of turbulence terms from the equation for $\overline{u}$ leads  to an ODE with solution resembling the `true' (recovered from DNS simulations) mean velocity profile with fine accuracy.
A work related to this study is found in paper~\cite{kovalishin2008transformation}, where the Navier--Stokes equations are given in angular variables. The author is unaware of any other literature, where the $u$--$\br$ variables were used to describe fluid dynamics.

\section{Speed--direction flow variables}~\label{sNSE}
Consider a  flow of incompressible Newtonian fluid in $\mathbb{R}^3$  governed by the Navier--Stokes  equations
\begin{equation}\label{NSE}
\left\{
\begin{aligned}
\frac{\partial \bu}{\partial t}+(\bu\cdot \nabla)\bu - \nu (\Delta \bu) +\nabla p&=\blf \\
\Div \bu&=0
\end{aligned}\right.
\end{equation}
with fluid velocity $\bu$, kinematic pressure $p$, kinematic viscosity coefficient $\nu>0$ and given body forces $\blf$.

We are interested in representing the fluid velocity $\bu(\bx,t)$ at point $\bx\in\mathbb{R}^3$, $t\ge0$,   in terms of its speed $u(\bx,t)\in\mathbb{R}$ and direction $\br(\bx,t)\in S^2$:
\begin{equation*}
\mathbf{u}=u \mathbf{r},\quad |\mathbf{r}|=1,\quad u\ge 0.
\end{equation*}
 For $\mathbf{u}=0$, the choice of $\mathbf{r}$ is not unique. Operators of vector calculus in new variables take the following form:
\begin{equation}\label{id1}
\begin{split}
% \nonumber % Remove numbering (before each equation)
  \nabla\bu& = u\nabla\br+\br\otimes(\nabla u),  \\
   \nabla\times\bu& = u(\nabla\times\br)+(\nabla u)\times\br, \\
  \Div\bu &= (\br\cdot \nabla u)+u \Div \br.
  \end{split}
\end{equation}
Diffusion and inertia terms are easily computed to be
\begin{equation}\label{id2}
% \nonumber % Remove numbering (before each equation)
\begin{split}
  \Delta\bu& = u\Delta\br+\br\Delta u+ 2[\nabla \br]\nabla u, \\
   (\bu\cdot\nabla \bu)& = u^{2}(\br\cdot \nabla)\br+u(\br \otimes \br) \nabla u.
%\\
%  (\nabla\times\bu)\times\bu &=& u^2(\nabla\times\br)\times\br+u((\nabla u)\times\br)\times\br\\  &=& u^2(\nabla\times\br)\times\br-u\bP(\nabla u).
\end{split}
\end{equation}
One finds several useful identities by differentiating equality $|\br|^2=1$ in time and space:
\begin{equation}\label{br-ident}
\frac{\partial \br}{\partial t}\cdot\br=0,\quad [\nabla \br]^T\br=0, \quad -\br\cdot\Delta\br=|\nabla\br|^2,
\end{equation}
where $|A|^2=\mbox{tr}(A^TA)$. From the second equality we also get
\begin{equation}\label{id3}
(\br\cdot \nabla)\br=2\bD(\br)\br=(\nabla\times\br)\times\br
\end{equation}
where $\bD(\br)=\frac12(\nabla\br+[\nabla\br]^T)$.

%For certain calculations, it is convenient to used the rotation form the inertial terms:
%\[
%(\bu\cdot \nabla)\bu+\nabla p=(\nabla\times\bu)\times\bu+\nabla P,
%\]
%with Bernoulli pressure $P=p+\frac12|\bu|^2$.

\subsubsection{Equations in new variables and the split system }
Thanks to the identities \eqref{id1}--\eqref{br-ident} the Navier--Stokes equations \eqref{NSE} in the speed--direction variables  take the form
\begin{equation}\label{NSEphth}
\left\{
\begin{aligned}
\br \frac{\partial u}{\partial t}+u \frac{\partial \br}{\partial t}+u^{2}(\br\cdot \nabla)\br+u(\br \otimes \br) \nabla u\qquad&\\ %+u^2(\nabla\times\br)\times\br-u\bP(\nabla u)
- \nu\left( \br\Delta u  +2 \left(\nabla u\cdot \nabla\right)\br + u \Delta \br\right)+\nabla p&=\blf \\
\br\cdot \nabla u+u \operatorname{div} \br&=0
\end{aligned}\right.
\end{equation}
We now split the  momentum equation by projecting it on the velocity direction and orthogonal plane. To this end, we take the scalar product of the momentum equation with $\br$ and note that due to \eqref{br-ident} and \eqref{id3} we have
\[
\begin{split}
\br\cdot\left(\nabla u\cdot \nabla\right)\br=\br^T[\nabla\br]\nabla u =(\nabla u)\cdot([\nabla\br]^T\br)=0\\
 u^{2}\br\cdot(\br\cdot \nabla)\br+u\br\cdot(\br \otimes \br) \nabla u =\br\cdot((\nabla\times\br)\times\br)+u\br\cdot\nabla u\\=
 {\small \frac12}\br\cdot\nabla u^2.
\end{split}
\]
 Employing this identities together with $|\br|^2=1$ and  $\frac{\partial \br}{\partial t}\cdot\br=0$, $-\br\cdot\Delta\br=|\nabla\br|^2$
 from \eqref{br-ident} we obtain the first equation of the split system
\begin{align}\label{NSEsplit1}
\frac{\partial u}{\partial t} - \nu \Delta u +\nu |\nabla\br|^2 u +\br\cdot\nabla (p+\frac{u^2}{2})&=\blf_r
\end{align}
with $\blf_r=\br\cdot\blf$.
We now project the momentum equation on the orthogonal planes to the fluid velocity directions by multiplying the first equation in \eqref{NSEphth} by the orthogonal projector $\bP=\bI-\br\otimes\br$ and use $\bP\br=0$. For the treatment of the viscous term, we also compute using \eqref{br-ident}:
\[
\bP\Delta\br=\Delta\br-(\br\cdot\Delta\br)\br=\Delta\br+|\nabla\br|^2\br.
\]
We arrive at the second equation of the split system:
\begin{equation}
\begin{split}
u \frac{\partial \br}{\partial t}+u^2(\br\cdot \nabla)\br
-2 \nu\left(\nabla u\cdot \nabla\right)\br\qquad&\\ -\nu u(\Delta \br+|\nabla\br|^2\br)+\bP\nabla p&=\blf_p,
\end{split}
\label{NSEsplit2}
\end{equation}
with $\blf_p=\bP\blf$. Alternative forms of \eqref{NSEsplit2} can be derived using \eqref{id3} and other expressions for the viscous terms.

\subsubsection{Energy balance}
With the help of \eqref{id1} and \eqref{br-ident} we find for the kinetic energy and diffusion densities:
\begin{equation}
  \frac12|\bu|^2=  \frac12u^2, \quad |\nabla\bu|^2 = u^2|\nabla\br|^2+|\nabla u|^2. \label{grad_id}
%   |\nabla\times\bu|^2 &=& u^2|\nabla\times\br|^2+|\bP\nabla u|^2+u(\nabla u)^T[\nabla\br]\br, \notag \\
%  (\nabla\times\bu)\cdot\bu &=& u^2(\nabla\times\br)\cdot\br. \notag
\end{equation}
For the energy balance, let us assume a flow in a finite volume $\Omega$ with no-slip  boundary condition $\bu=0$ on the boundary $\partial\Omega$.
Then for any smooth solution to the Navier--Stokes equation the energy equality follows by multiplying  \eqref{NSEsplit1} by $u$, integrating over  $\Omega$ and invoking the continuity equation:
\begin{equation*}
\frac12\frac{d}{dt}\int_\Omega u^{2}\,dx+\nu\int_\Omega\left( |\nabla u|^{2}+\left|\nabla \br\right|^2 u^2\right)\,dx=\int_\Omega \blf_r u\,dx.
\end{equation*}

We see that the information about the energy balance is essentially  encoded by the equation \eqref{NSEsplit1} and the continuity equation.
This motivates our focus on  \eqref{NSEsplit1}, when we are interested in a possible role of the speed--direction decomposition in understanding turbulent flows.

\section{Turbulent variables}
\begin{figure*}
\begin{center}
 \includegraphics[width=0.45\textwidth]{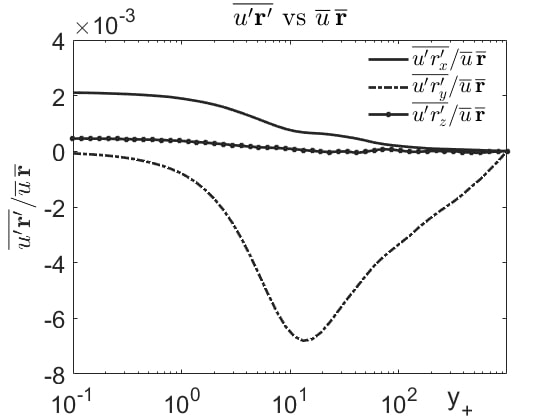}\quad
 \includegraphics[width=0.45\textwidth]{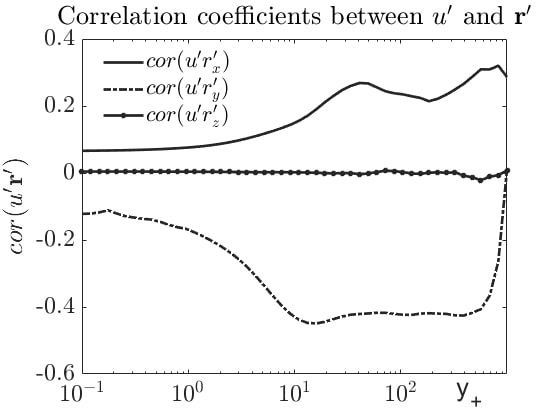}
  \end{center}
 \caption{Left: The relative magnitude of correlation vector $\overline{u'\br'}$ with respect to mean profile $\overline{u}\,\overline{\br}$.
Right: Correlation coefficients for  $u'$ and $\br'$  vs.  the distance from the channel wall in viscous units.}\label{fig:cov_cor}
\end{figure*}
Further $\overline{f}$ denotes an ensemble  average of quantity $f$ so that  $f^\prime=f-\overline{f}$ is a turbulent fluctuation. For the mean flow velocity it holds
\begin{equation}\label{v_decomp}
\overline{\bu}=\overline{u\br}= \overline{u}\,\overline{\br}+ \overline{u'\br'}.
\end{equation}
If fluctuations in flow direction and speed are linearly independent (uncorrelated) statistics, then  \eqref{v_decomp} simplifies to
\begin{equation}\label{iso_decomp}
\overline{\bu}= \overline{u}\,\overline{\br}.
\end{equation}
The assumption leading to \eqref{iso_decomp} holds for {isotropic} turbulence. In general, isotropy  is a stronger
%\begin{wrapfigure}{r}{8cm}
%\vspace{-20pt}
\begin{figure*}
\begin{center}
 \includegraphics[width=0.45\linewidth]{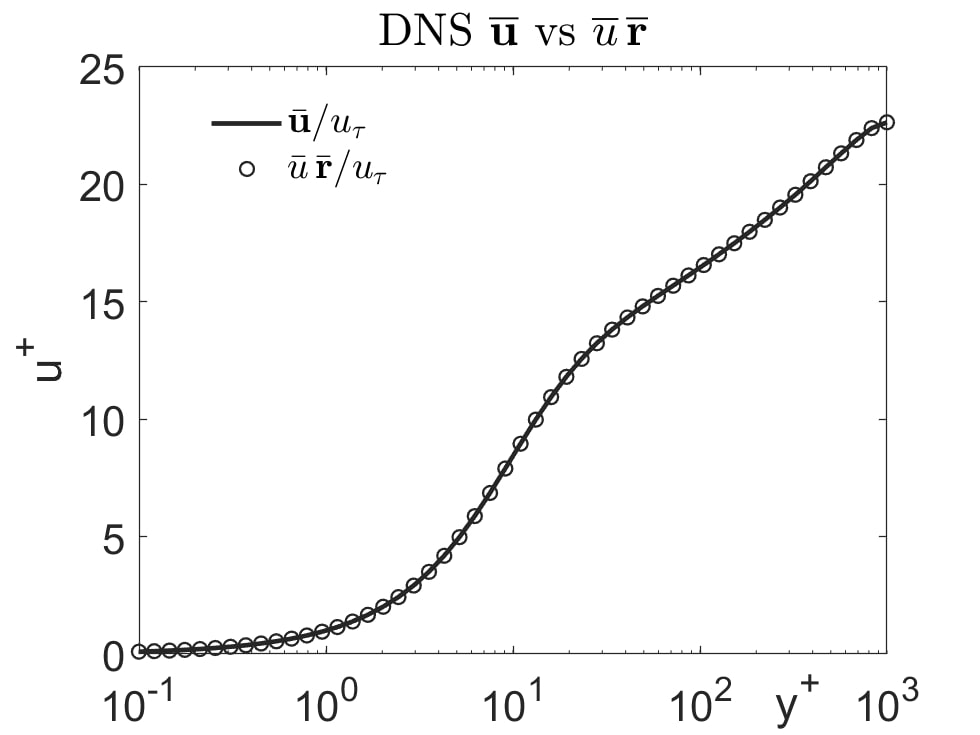}\quad
 \includegraphics[width=0.45\textwidth]{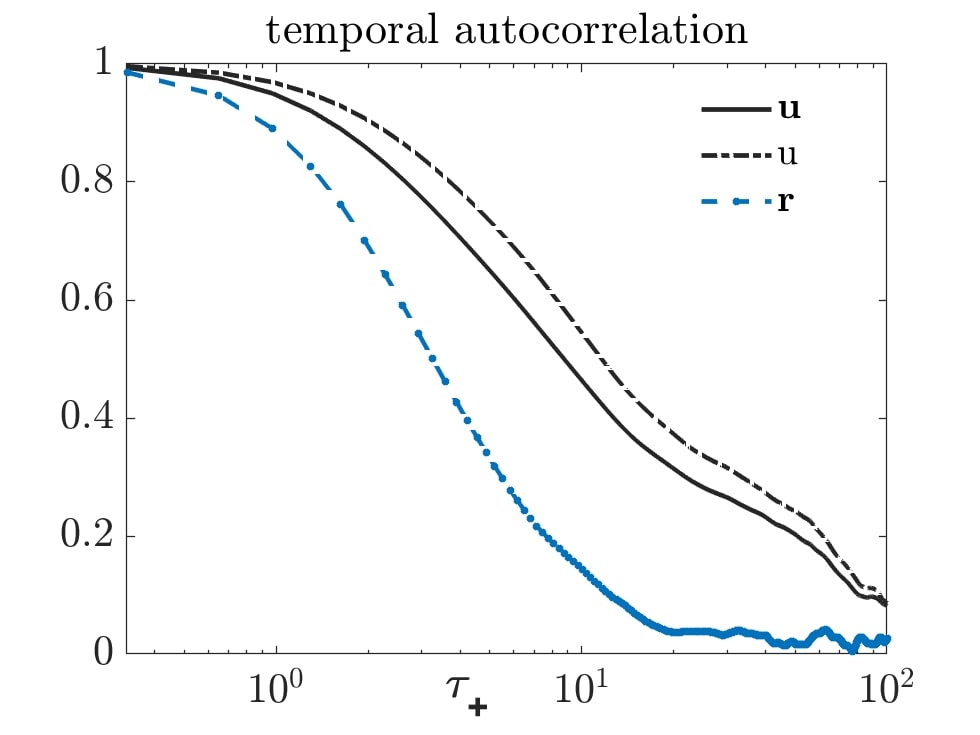}
  \caption{\small Left: Mean velocity profiles $\overline{\bu}$ and $\overline{u}\,\overline{\br}$ in viscous units.
  Right: Temporal auto-correlation  in $\bu$ (matrix norm), $u$ and $\br$ (matrix norm) vs. viscous time at $y^{+}=153$. Same faster decay in ACF for $u$ is observed for all $y^{+}>0$ across the channel.  }\label{fig:profiles}
\end{center}
\end{figure*}
%\vspace{-20pt}
%\end{wrapfigure}
assumption then $\overline{u'\br'}=0$.
While for anisotropic turbulence we do not see a reason for $u'$ and $\br'$ to be independent, we  hypothesize that for many flows with correlated fluctuations in speed and direction \textit{the values of $\overline{u'\br'}$ are   small relative
to the mean flow and \eqref{iso_decomp} holds approximately}. To back this hypothesis, we { first} evaluate the statistics of interest using the data set from JHU Turbulence Databases~\cite{kanov2015johns}  for the { pressure gradient driven} turbulent channel flow with $Re_\tau=10^3$ (bulk Reynolds number $Re=4\times10^{4}$, viscous length scale equals $10^{3}$, friction velocity $u_\tau=4.9968\times10^{-2}$).\footnote{Data was sampled for 50  distances from the wall, which  were equality distributed  in the log scale for  $y_+\in[0.1,10^{3}]$. For each $y_+$,  the data (velocity, velocity gradient, and pressure gradient) were collected over 961 points uniformly distributed in the $xz$-plane and for 1000 time instances uniformly distributed in the simulation time interval $[0,26]$.  Ensemble means for each $y_+$ node were computed as averages over the data in $961\times 1000$ space--time points.}
From the left plot in Figure~\ref{fig:cov_cor} we see that
the relative  correlations $|\overline{u'\br'}|/|\overline{u}\,\overline{\br}|$ are of order $10^{-3}$ for all distances from the channel wall, which suggests that  \eqref{iso_decomp} holds with excellent accuracy.   As a result, the  mean flow $\overline{\bu}$ and
$\overline{u}\,\overline{\br}$ are in agreement as illustrated in Figure~\ref{fig:profiles} (left), where all three averaged statistics were
computed from the database DNS results: The computed $\overline{\bu}_x$ and $\overline{u}\,\overline{\br}_x$ virtually coincide. %For the reference purpose, the plot shows the linear law and log-law curves.
We note that for the shear turbulence (such as the channel turbulence) $u'$ and $\br'$ are not necessary  uncorrelated, and the right plot in Fig.~\ref{fig:cov_cor} shows that the correlation is not insignificant between $u'$ and $x$, $y$ components of $\br'$. Nevertheless, it turns out to be reasonable to accept \eqref{iso_decomp}. { The experiment was repeated for the data acquired from the JHU database for the turbulent channel flow with the higher friction Reynolds number of $Re_\tau=5200$, and results show very similar  relative  correlations $|\overline{u'\br'}|/|\overline{u}\,\overline{\br}|$ of order $10^{-3}$ (not visualized here).}

In addition, the right plot in Figure~\ref{fig:profiles} shows normalized temporal autocorrelation functions (ACF) for $\bu$, $u$ and $\br$, where for the vector quantity ACF is defined as the Frobenius norm of the autocorrelation matrix\footnote{For computing AFC we sample data  in  30 points randomly distributed in the $xz$-plane and in 4000 time instances uniformly distributed in the simulation time interval $[0,26]$}. We see that the flow fluctuations in a point tend to forget their directions faster than the speed. This may be one factor explaining weaker correlation between $\br'$ and $u'$. The ACF are shown for   $y^{+}=153$, but similar picture is observed for other distances.
{ A further insight can be gained from an estimate of  $|\overline{u'\br'}|$ through turbulence intensities and correlation coefficients. Let $u_{rsm}=\left(\overline{|u'|^2}\right)^{\frac12}$ denote the root-mean-square (rms) of fluctuations in $u$. Similar definition applies to vector quantities $\bu$, $\br$ and their components (we use $r_{x_1},\,r_{x_2},\,r_{x_3}$ notation below for Cartesian components of vector $\br$, and similar for $\bu$). %, e.g.$u_i=\bu\cdot\be_i$.
The turbulent intensity in $u$ is the ratio of fluctuations rms to the mean flow speed,
$I_{u}=u_{rsm}/\overline{u}$, and similarly for each $\br$ component $I_{r_i}=r_{x_i,rsm}/|\overline{\br}|$, {\small $i=1,2,3$}. The total turbulent intensity in $\br$ is than $I_{\br}=(\sum_{i=1}^{3}I_{r_i}^2)^{\frac12}=\br_{rsm}/|\overline{\br}|$.
One can write $|\overline{u'\br'}|$ relative to the mean flow in terms of correlation coefficients $c_{x_i}$ (these are coefficients depicted in  Fig.~\ref{fig:cov_cor}, right plot)  and intensities:
\begin{equation}\label{aux233}
\frac{|\overline{u'\br'}|}{|\overline{u}\,\overline{\br}|}=I_u\Big(\sum_{i=1}^3  c_{x_i}^2 I_{r_i}^2\Big)^{\frac12}\le \max_{i}|c_{x_i}|I_uI_{\br}.
\end{equation}
We are now interested in estimating $I_u$ and $I_{r_i}$ (or $I_{\br}$) in terms of turbulent intensities in the velocity field, $I_{u_i}=u_{x_i,rsm}/|\overline{\bu}|$, {\small $i=1,2,3$},   the statistics commonly reported in the literature. From the identities $u_{rsm}^2=\overline{|u'|^2}=\overline{|\bu|^2}-\overline{|\bu|}^2$ and $\bu_{rsm}^2=\overline{|\bu'|^2}=\overline{|\bu|^2}-|\overline{\bu}|^2$ and $|\overline{\bu}|\le\overline{|\bu|}$ we conclude that
\[
I_u\le I_{\bu},\quad\text{with}~I_{\bu}=(\sum_{i=1}^{3}I_{u_i}^2)^{\frac12}=\bu_{rsm}/|\overline{\bu}|.
\]
%Obviously, we have $u_{rsm}\le I_u \overline{u}$.
For the turbulent intensities in $\br$  it holds
\begin{equation}\label{aux239}
I_{r_i}\le (I_{u_i}+(1+|c_{x_i}|)I_u)|\overline{\br}|^{-1}.
\end{equation}
This estimate follows from the decomposition (it is easy to check using \eqref{v_decomp} componentwise)
\[
r_{x_i}'=\left( u_{x_i}'+\overline{u'r_{x_i}'}-u'r_{x_i}\right)/\overline{u}
\]
and the triangle inequality, which gives
\[
\begin{split}
r_{x_i,rms}&\le \left(u_{x_i,rms}+|\overline{u'r_{x_i}'}|+\big(\overline{|u'r_{x_i}|^2}\big)^{\frac12}\right)/\overline{u} \\
&\le \left(u_{x_i,rms}+ |c_{x_i}|u_{rsm} +u_{rsm}\right)/\overline{u}
\end{split}
\]
where for the last estimate we use the Cauchy-Schwarz inequality and $r_{x_i,rms}\le1$, $|r_{x_i}|\le1$.
Equations \eqref{aux233}, \eqref{aux239} and $u_{rsm}\le I_u \overline{u}$ imply the following bound:
\begin{equation}\label{aux242}
\frac{|\overline{u'\br'}|}{|\overline{u}\,\overline{\br}|} \le I_{u}\Big(\sum_{i=1}^3  c_{x_i}^2 (I_{u_i}+(1+|c_{x_i}|)I_u)^2\Big)^{\frac12}|\overline{\br}|^{-1},
\end{equation}
or making the rough estimates $|c_{x_i}|\le1$, $I_{u}\le I_{\bu}$, $I_{u_i}\le I_{\bu}$, we have
\[
|\overline{u'\br'}|\le 2\sqrt{5}\,I_{\bu}^2\,\overline{u}.
\]

\begin{figure}
\begin{center}
 \includegraphics[width=0.45\textwidth]{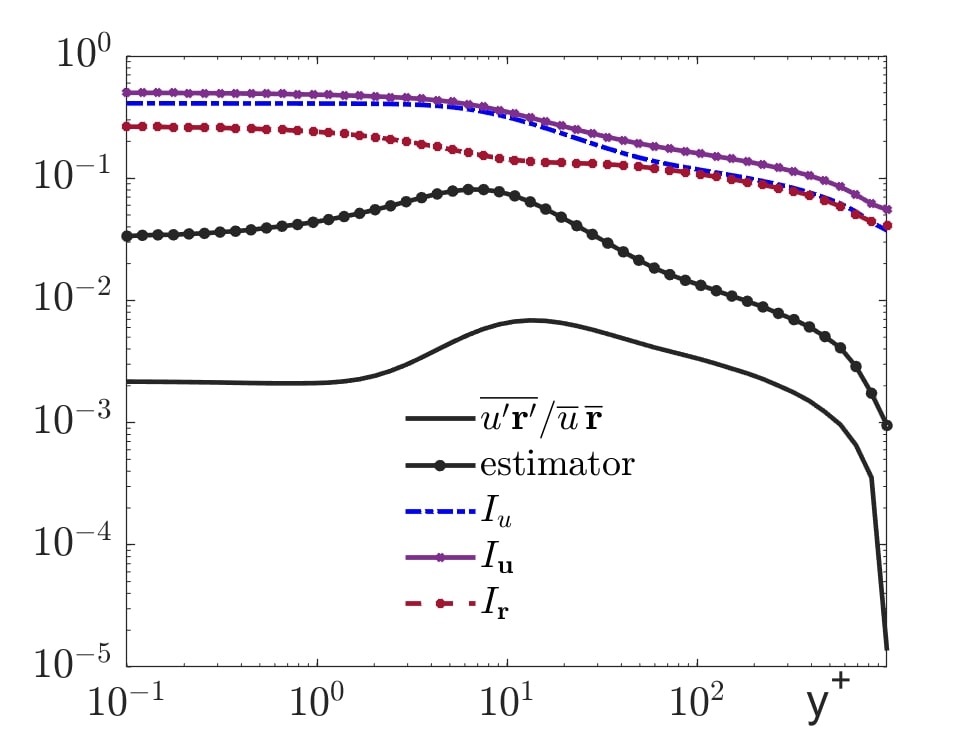}
  \end{center}
 \caption{Relative norm of $\overline{u'\br'}$ vs estimator from \eqref{aux242}.  The plot also shows turbulent intensities in $u$, $\bu$, and $\br$. }\label{quant}
\end{figure}
Thus the  correlation part in \eqref{v_decomp} decreases \emph{quadratically}  with the turbulent intensity of the velocity field. Note that due to $\br_{rms}\le1$, the linear estimate
$|\overline{u'\br'}|\le I_u\overline{u}$ is trivial.  The quantities involved in \eqref{aux242} are illustrated in Fig.~\ref{quant}: the intensities $I_u$, $I_\bu$, and $I_{\bf r}$ in the main stream vary between 0.05 and 0.2. In this example $I_{u}\le I_{\bu}$ and $I_{\br}\le I_{\bu}$ hold for all $y_{+}$.  The ``estimator'' denotes the graph of the term on the right-hand side of \eqref{aux242}.
In general, the estimator graph follows $|\overline{u'\br'}|/|\overline{u}\,\overline{\br}|$, but overestimates it quite significantly, suggesting that  \eqref{aux242} can be still nonoptimal. %In particular, in this example turbulent intencities in $\br$ appear to be smaller than in $\bu$; cf. Fig.~\ref{quant}.
Again, same observations were made for $Re_\tau=5200$ channel flow data (not visualized here).
}

Motivated by the above analysis and observations, we assume   \eqref{iso_decomp}.  In this case, we  see that $\overline{u}$ determines the mean flow $\overline{\bu}$ once the mean direction $\overline{\br}$ is known with a reasonable certainty (as in the case of the turbulent channel flow).
Therefore, it is interesting to look at  the  equation for $\overline{u}$. We call $ \overline{u}$ \textit{the mean flow profile}.
Taking the ensemble average of \eqref{NSEsplit1} we arrive at
\[
\frac{\partial \overline{u}}{\partial t} - \nu\left(\, \Delta \overline{u} - \overline{|\nabla\br|^2 u}\right) +\overline{\br\cdot\nabla P}=\overline{\blf_r},\quad\text{with}~P=p+\frac{u^2}{2},
\]
or working out the averages we get
\begin{equation}\label{mean_prof}
\frac{\partial \overline{u}}{\partial t} - \nu\left( \Delta \overline{u} - \overline{|\nabla\br|^2}\overline{u} - \mathcal{Q}
\right)
+\overline{\br}\cdot\nabla(\frac12\overline{u}^2+\overline{p})+\mathcal{P}=\overline{\blf_r},
\end{equation}
with  turbulent correlation functions:
\[
\mathcal{Q}=\overline{(|\nabla\br|^2)' u' }\quad\text{and}\quad \mathcal{P}=\overline{\br'\cdot\nabla P'}+{\small\frac12}\overline{\br}\cdot\nabla\overline{|u'|^2}.
\]
%We see that %unlike the equation governing the  mean flow $\overline{\bu}$,
  The full Reynolds turbulent stress tensor does not appear in \eqref{mean_prof} and the action of fluctuations on the mean flow profile comes through the  total pressure correlation with $\br'$, variation of $\overline{|u'|^2}$ along mean flow directions, extra  viscous terms $\mathcal{Q}$, and $\overline{|\nabla\br|^2}$ factor in front of $\overline{u}$. Let us take a closer look at  these terms for the channel turbulence example, where they all are functions of $y$.

\begin{figure*}
\begin{center}
 \includegraphics[width=0.45\textwidth]{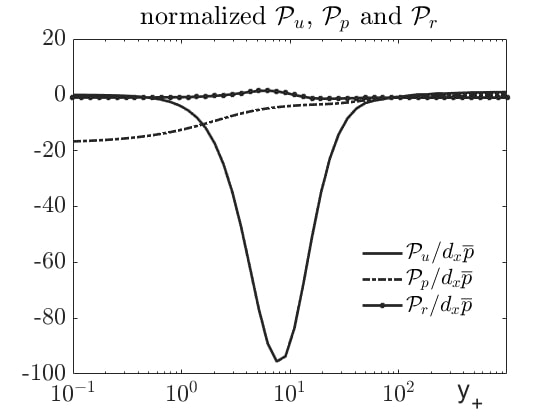}\quad
  \includegraphics[width=0.45\textwidth]{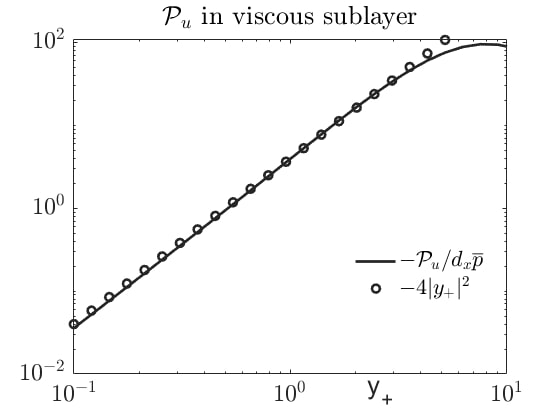}
  \end{center}
 \caption{Left: Turbulent functions $\mathcal{P}_u$, $\mathcal{P}_p$ and $\mathcal{P}_r$ normalized by   $\partial_x\overline{p}$ vs. the distance from the channel wall in viscous units. Right: Fitting  of $\mathcal{P}_u$ normalized by   $\partial_x\overline{p}$ by a quadratic function in the viscous sublayer. }\label{fig:cov_pe}
\end{figure*}
 We start with splitting   $\mathcal{P}$ into the parts corresponding to the mean variation of turbulent kinetic energy and pressure along $\br'$ and the  variation of $\overline{|u'|^2}$ along $\overline{\br}$:
 \[
  \mathcal{P}=\underbrace{{\small\frac12}\overline{\br'\cdot\nabla (u^2)'}}_{\mathcal{P}_u}~+\underbrace{\phantom{\small\frac12}\overline{\br'\cdot\nabla p'}}_{\mathcal{P}_p}~ +~\underbrace{{\small\frac12}\overline{\br}\cdot\nabla\overline{|u'|^2}}_{\mathcal{P}_r}.
 \]
Function $\mathcal{P}_u$ has a meaning of a correlation between  fluctuations in the flow direction and fluctuations in the kinetic energy gradient. Far from the wall, the mean kinetic energy has little variation in space so that on \textit{average} a fluctuation in the flow direction should not cause much of energy flux. Hence we expect  $\mathcal{P}_u$ to be small there. In the laminar sublayer  $\mathcal{P}_u$ is small for a different reason, namely the turbulent fluctuations are insignificant there. The situation differs in the transition (overlap)
region between the viscous sublayer and the outer region. In this region,   fluctuations in $\br$ are significant and they produce the mixing which transfers the energy from the turbulent stream to the viscous layer, where it dissipates. This scenario explains the behaviour of   $\mathcal{P}_u$ recovered from the turbulent data\footnote{Note that $\nabla u$, $\nabla (u^2)$ and $\nabla\br$ are not available directly in the database and we recover them using the (available) velocity gradient  through the equalities
\[
\nabla u=\br^T[\nabla\bu],\quad \nabla (u^2)=2u\nabla u, \quad\text{and}\quad|\nabla\br|^2=\frac{|\bP\nabla\bu|^2}{u^2}.
\]
The first identity follows from \eqref{id1} and \eqref{br-ident}. In turn, \eqref{grad_id} and $\nabla u=\br^T[\nabla\bu]$ implies the expression for $|\nabla\br|^2$.}
 and shown in Figure~\ref{fig:cov_pe} (left), where we see a strong negative correlation reflected by a log-Gaussian type peak around $y_+=8$. In a viscous sublayer, we observe a clear $\mathcal{P}_u\simeq 4y_+^2$ asymptotic, while understanding the  asymptotic of decay  for $y_+\to+\infty$ needs further insights.
The same left plot in  Figure~\ref{fig:cov_pe} shows $\mathcal{P}_p$ and $\mathcal{P}_r$ (all quantities are normalized by the mean pressure drop). These functions play minor role in the transition region, but knowing their asymptotics for   $y_+\to+\infty$ is important for the correct prediction of $\overline{u}$ as we see later.

\begin{figure*}
\begin{center}
 \includegraphics[width=0.45\textwidth]{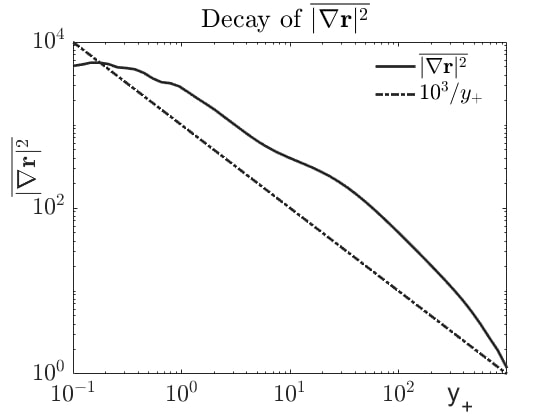}\quad
  \includegraphics[width=0.45\textwidth]{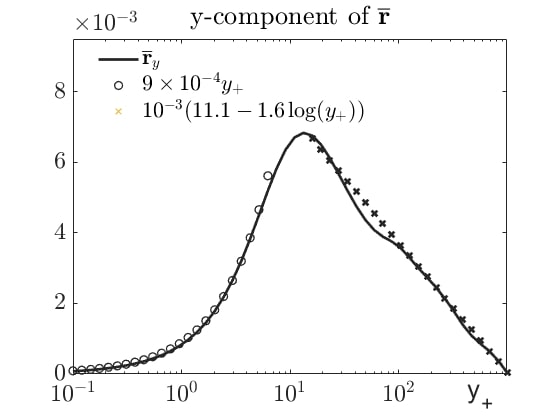}
  \end{center}
 \caption{The mean  squared norm of the flow direction gradient (right) and the $y$-component of the mean flow direction $\overline{\br}$  (left) vs the distance from the channel wall in viscous units.}\label{fig:grad_e}
\end{figure*}

To get a rough idea about the $y$-dependence of  $\overline{|\nabla\br|^2}$, %the mean squared variation of the direction coefficient
 it can be useful to adopt the view of turbulent flow as a hierarchy of vortices  so that the flow on a distance
 $y$ from the wall is dominated (in an average sense) by  vortices of size $O(y)$. The center of  a standing 2D vortex is a singular point of $\nabla\br$ so that the integral of $|\nabla\br|^2$ diverges logarithmically. The situation is more complicated in 3D, but it looks reasonable to suggests that  $\overline{|\nabla\br|^2}(y)$ growth proportional to the average number of $O(y)$-vortices filling the layer. Hence we may expect $\overline{|\nabla\br|^2}$ to increase  closer to the wall and decrease in the developed turbulent stream. The plot of   $\overline{|\nabla\br|^2}$ reconstructed from the DNS data in Figure~\ref{fig:grad_e} (left) confirms this hypotheses and shows that $\overline{|\nabla\br|^2}=O(y^{-1})$ can be a reasonable (though not perfect) approximation.

 The derivative of the mean-flow  kinetic energy along the  mean direction $\overline{\br}$ plays also an important
role  in \eqref{mean_prof}. Figure~\ref{fig:grad_e} reveals that although the deviation of $\overline{\br}$ from the $x$-direction is less then 1\%, the $y$-component  of  $\overline{\br}$ demonstrates quite distinctive behaviour, first growing in the viscous layer close to $\overline{\br}(y)\approx y=10^{-3}y_+$ and in the transition region changing this growth to a slow decay, with the  log function turning out to be a very  good fit {(similarity of graphs in figures~\ref{fig:cov_cor}(left) and~\ref{fig:grad_e}(right) is not coincidence, since the quantities are related through \eqref{v_decomp})}. Finally, $\nu Q$ is small due to the scaling with viscosity coefficient and can be neglected in the equation for $\overline{u}$.

\begin{figure}
\begin{center}
 \includegraphics[width=0.45\textwidth]{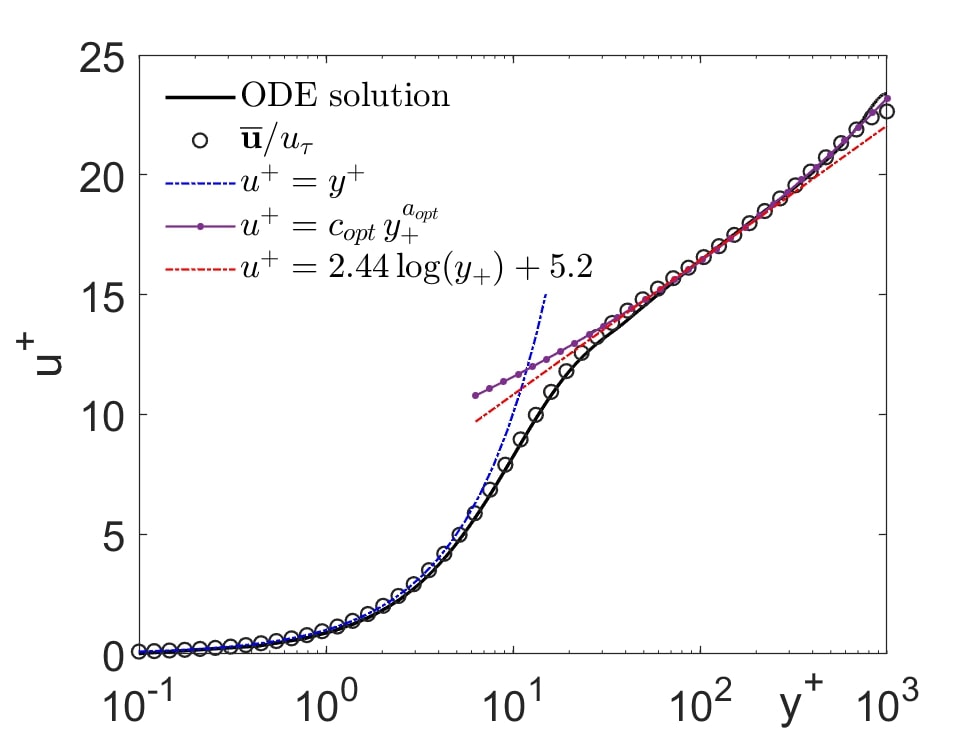}
\end{center}
 \caption{Re-scaled by $u_\tau^{-1}$ solution to \eqref{ODE} with data-reconstructed coefficients vs computed $\overline{\bf u}$. }\label{fig:ProfSol}
\end{figure}

For the example of turbulent channel flow we assume the statistically stationary turbulence with $\overline{u}$ and coefficients in \eqref{mean_prof} independent of $x$ and $z$. The equation reduces to
\begin{equation}\label{ODE}
 - \nu\left(\frac{d^2\overline{u}}{dy^2} - \overline{|\nabla\br|^2}\overline{u}\right)
+\overline{\br}_y\overline{u}\frac{d\overline{u}}{dy}+\mathcal{P}=-\overline{\br}\cdot\nabla\overline{p},
\end{equation}
%One computes that $\overline{|\nabla\br|^2}=|\nabla\overline{\br}|^2+\overline{|\nabla\br'|^2}$. The first term on the right hand side in this decomposition for  $\overline{|\nabla\br|^2}$ is insignificant compared to the second one.
Based on the above discussion we use the following analytical representations for the statistics appearing
 in   \eqref{ODE}:
 \begin{widetext}
\begin{equation}\label{Coeff}
\begin{aligned}
 \overline{|\nabla\br|^2}&\approx r_g(y_+)=3\times10^{-3} y_+^{-1},\quad Q=0,\\
\overline{\br}_y&\approx r_y(y_+)=\left\{
\begin{array}{ll}
10^{-3}y_{+}&~\text{for}~y_{+}\le 8\\
10^{-3}(11.1-1.6\ln y_+)&~\text{for}~y_{+}> 8
\end{array}
\right.,\\
 \mathcal{P}+\overline{\br}\cdot\nabla\overline{p}&\approx r_p(y_+)
 =\left\{
\begin{array}{ll}
c_0&~\text{for}~y_{+}\le 3\\
c_1\exp\left( - \frac{(\ln(y_+)-\ln(\widehat{y_+}))^2}{2\sigma^2} \right)&~\text{for}~3\le y_{+}\le 40\\
c_2 y_+^{\alpha}&~\text{for}~y_{+}> 40
\end{array}
\right.
\end{aligned}
\end{equation}
\end{widetext}
The solution of \eqref{ODE} was not sensitive to $c_0$ and we set $c_0=0$, while the correct position and amplitude of
the log-Gaussian and the proper decay of $\mathcal{P}$ in the interior were found to be both important for   \eqref{ODE} to correctly reproduce the `true' mean profile. We set $c_1=-0.25$, $\widehat{y_+}=7.9$, $\sigma=0.66$ and $c_2=-2.75$, $\alpha=1.33$. The solution to \eqref{ODE} with the coefficients defined in \eqref{Coeff} and boundary conditions $\overline{u}(0)=0$, $d_y\overline{u}(1)=0$  is shown in Figure~\ref{fig:ProfSol}. { It appears close to  the mean profile recovered from the DNS data, confirming the validity of approximations in \eqref{Coeff}. For the reference purpose we also include in Figure~\ref{fig:ProfSol} the log-law of the wall with the standard choice of coefficients and the power law. Since for the power law, $\overline{u}_{+}=cy_+^{\alpha}$ there is no one accepted choice of $\alpha$ and $c$, see, e.g. the discussion in~\cite{buschmann2006recent,egolf2020first}, we use the values of $c_{opt}=8.17$ and $\alpha_{opt}=0.151$  that have been found by the least square fit of the power curve to the DNS data for $y_{+}\ge25$. Both laws  give reasonably good prediction of the mean flow profile in the turbulent stream for $y_{+}\ge30$. The ODE  solution have a slight preference towards  the power law  for larger $y_+$, but also recovers the correct profile in the laminar and transition zones. }

%\begin{wrapfigure}{r}{8cm}
%\vspace{-20pt}
%\begin{center}
%\end{center}
%\vspace{-20pt}
%\end{wrapfigure}

\subsection{Energy balance}
We comment on the energy balance for the averaged turbulent flow in the speed--direction variables.
Similar to the mean flow velocity, mean profile can be related to the  kinetic energy $\overline{E}$:
\begin{equation}\label{energy_decomp}
 2\overline{E}=\overline{u^2}=\overline{u}^2+\overline{|u^\prime|^2}.
\end{equation}
The decomposition \eqref{energy_decomp} is, in general, different from
$2\overline{E}= \overline{|{\bu}|^2}= |\overline{\bu}|^2 + \overline{|\bu^\prime|^2}$,
where the evolution of the mean flow energy $\frac12 |\overline{\bu}|^2$ is driven by viscous tensions and Reynold's stress. Full Reynold's stress does not appear in equation for $\overline{u}^2$, which we get by multiplying \eqref{mean_prof} with $\overline{u}$:
\begin{equation*}%\label{mean_erg}
\begin{split}
\frac12\frac{\partial\, \overline{u}^2}{\partial t} - \nu (\Delta \overline{u})\overline{u} +\nu\left(\overline{|\nabla\br|^2} \overline{u}^2 +
 Q \overline{u} \right)&\\ +\overline{\bu}\cdot\nabla\overline{P}-\overline{u'\br'}\cdot\nabla\overline{P}+\overline{u}\mathcal{P}&=\overline{\blf_r}\overline{u},\quad
 \overline{P}=\frac12\overline{u}^2+\overline{p},
 \end{split}
\end{equation*}
where we used \eqref{v_decomp}.  Again under the assumption about a weak correlation between speed and direction in fluctuations the term $\overline{u'\br'}\cdot\nabla\overline{P}$ can be omitted and the above identity simplifies to
\begin{equation}\label{mean_erg}
\frac12\frac{\partial\,\overline{u}^2}{\partial t} - \nu (\Delta \overline{u})\overline{u} +\nu\left(\overline{|\nabla\br|^2} \overline{u}^2 +
 Q \overline{u} \right) +\overline{\bu}\cdot\nabla\overline{P}+\overline{u}\mathcal{P}=\overline{\blf_r}\overline{u},
\end{equation}
Denote by $\frac{\rm \overline{d}\cdot}{\rm dt}=\frac{d\cdot}{dt}+(\overline{\bu}\cdot\nabla)\cdot$ the material
derivative along mean flow trajectories. One can regroup terms as
\[
\frac12\frac{\partial\,\overline{u}^2}{\partial t} + \overline{\bu}\cdot\nabla\overline{P}+\overline{u}\mathcal{P}=
\frac12{\frac{{\rm\overline{d}}\, \overline{u}^2}{\rm dt}}+  \overline{\bu}\cdot\nabla\widetilde{p}+\overline{u}(\mathcal{P}_u+\mathcal{P}_p),
\]
{with} $\widetilde{p}=\overline{p}+\frac12\overline{|u^\prime|^2}$.
Now let $\overline{V}$ be a material volume evolving with the mean flow field $\overline{\bu}$. Integrating the energy equation \eqref{mean_erg} over $\overline{V}$, using $\Div\overline{\bu}=0$ and the above relation we obtain:
\begin{widetext}
\begin{equation*}
\frac12\frac{\rm d}{\rm dt}\int_{\overline{V}} \overline{u}^2 =
 - \underbrace{\nu \int_{\overline{V}}\left\{ |\nabla \overline{u}|^2 +\overline{|\nabla\br|^2}\,\overline{u}^2 + Q \overline{u}\right\}}_{\text{viscous dissipation}}
  +\underbrace{\int_{\overline{\partial V}} \nu \overline{u}(\bn\cdot\nabla \overline{u}) + (\overline{\bu}\cdot\bn)\widetilde{p}}_{\text{energy flux on}~ \partial V}
  - \underbrace{\int_{\overline{V}}\overline{u}(\mathcal{P}_u+\mathcal{P}_p)}_{\text{work of}~ \mathcal{P}_u\&\mathcal{P}_p}
  +
  \int_{\overline{V}}\overline{\blf_r}\overline{u}.
\end{equation*}
\end{widetext}

We see that the effect of turbulent fluctuations on the mean flow energy balance is present through the work of correlation functions $\mathcal{P}_u$ and $\mathcal{P}_p$ and the boundary flux of $\overline{|u'|^2}$.

\section{Conclusions} The speed--direction variables have a potential to become a useful alternative approach to describe the motion
of fluids, and in particular of turbulent flows.
A mostly data-driven approach was taken here to understand  turbulent (correlation) functions arising in the mean profile equation, { i.e., the data from full numerical simulations was still used to define coefficient in the model \eqref{ODE}.}
More study is required to model them for more general flows {and understanding their dependence on flow parameters}. The paper does not discuss the complementing equation~\eqref{NSEsplit2}.
A suitable way to use it in modelling and analysis  has to be found.

\begin{acknowledgments}  Partial support from NSF through DMS-2011444 is acknowledged.    Matlab scripts for data sampling, postprocessing and ODE solution used to produce results reported in this paper are available from the author on reasonable request.
\end{acknowledgments}

\bibliography{literature}{}
\end{document}